\documentclass[aps,twocolumn,showpacs,groupedaddress]{revtex4}
\usepackage{amssymb,amsmath}
\usepackage{graphicx,array}
\usepackage{dcolumn}
\usepackage{bm}

\begin{document}
\title{Possible superfluidity of molecular hydrogen in a two-dimensional
       crystal phase of sodium}

\author{Claudio Cazorla}
\affiliation{Institut de Ci$\grave{e}$ncia de Materials de
             Barcelona (ICMAB-CSIC), 08193 Bellaterra, Spain}
\author{Jordi Boronat}
\affiliation{Departament de F\'{i}sica i Enginyeria Nuclear,
             Universitat Polit\`{e}cnica de Catalunya, Campus
             Nord B4-B5, E-08034, Barcelona, Spain}
\email{ccazorla@icmab.es}

\begin{abstract}
We theoretically investigate the ground-state properties of a  
molecular para-hydrogen (p-H$_{2}$) film in which crystallization is
energetically frustrated by embedding sodium (Na) atoms periodically distributed 
in a triangular lattice. In order to fully deal with the quantum nature 
of p-H$_{2}$ molecules, we employ the diffusion Monte Carlo method and 
realistic semi-empirical pairwise potentials describing the interactions 
between H$_{2}$-H$_{2}$ and Na-H$_{2}$ species. In particular, we calculate 
the energetic, structural and superfluid properties of two-dimensional 
Na-H$_{2}$ systems within a narrow density interval around equilibrium at 
zero temperature. In contrast to previous computational studies 
considering other alkali metal species such as rubidium and potassium, we find
that the p-H$_{2}$ ground-state is a liquid with a significantly large 
superfluid fraction of $\rho_{s}/\rho = 0.29(2)$. 
The appearance of p-H$_{2}$ superfluid response is due to the fact that the 
interactions between Na atoms and H$_{2}$ molecules are less attractive than 
between H$_{2}$ molecules. This induces a considerable reduction of the hydrogen 
density which favours the stabilization of the liquid phase. 
\end{abstract}

\pacs{67.70.+n, 67.90.+z, 61.50.Ah, 67.80.-s}

\maketitle

\section{Introduction}
\label{sec:intro}
Unlike helium, bulk molecular para-hydrogen (p-H$_{2}$) always solidifies if a 
sufficiently low temperature is reached~\cite{vanstraaten}. Intermolecular
H$_{2}$-H$_{2}$ interactions are attractive and quite intense hence, even 
though hydrogen molecules are lighter than $^{4}$He atoms, p-H$_{2}$ 
crystallization is energetically favoured over melting in the $T \to 0$ limit 
frustrating so any possibility to observe superfluidity (SF) or Bose-Einstein 
condensation (BEC) in bulk. Putting this into numbers, molecular hydrogen 
becomes a solid at temperatures below $T_{t} \sim 14$~K whereas the critical 
temperature at which BEC and SF are expected to occur is $T_{c} \sim 1$~K~\cite{boronat12}. 
In spite of that, many experimental attempts have focused on supercooling 
bulk liquid p-H$_{2}$ below $T_{c}$, although unfortunately with no apparent 
success to date~\cite{seidel,maris}.

A likely way to induce superfluidity in molecular hydrogen consists in lowering 
its melting temperature by reducing its dimensionality and/or confining it to restricted 
geometries. Following this line of thinking many experimental and theoretical studies 
have focused on the characterization and analysis of p-H$_{2}$ films adsorbed on  
different substrates~\cite{dekinder,brewer,schindler,sokol}. 
For instance, two-dimensional hydrogen has been observed to freeze at temperatures 
around $5$~K when placed onto an exfoliated graphite plate~\cite{liu}. Also, it has been 
experimentally shown that small para-hydrogen clusters immersed in $^{4}$He 
droplets exhibit superfluid-like behaviour~\cite{grebenev}. On the theoretical side, it 
has been predicted that one-dimensional arrays of p-H$_{2}$ molecules remain in the liquid 
phase down to absolute zero~\cite{boro} and that small two- and three-dimensional 
clusters of pure p-H$_{2}$ are superfluid at temperatures below $1-2$~K~\cite{gordillo01,drops}. 

An alternative way to induce superfluidity in molecular para-hydrogen 
may consist in embedding alkali metal (AM) atoms on it. This idea was originally 
proposed by Gordillo and Ceperley (GC)~\cite{gordillo} and is based on the fact that 
the interactions between alkali metal atoms and molecular hydrogen are less attractive than 
between p-H$_{2}$ molecules. Therefore, a substantial reduction of the 
equilibrium hydrogen density can be induced which triggers stabilization of the 
liquid. In particular, GC investigated two-dimensional AM-H$_{2}$ (AM = K and Cs) 
systems at low temperatures (i.e., $1-4$~K) employing the path integral Monte Carlo 
(PIMC) technique. They found that the p-H$_{2}$ equilibrium state in AM-H$_{2}$ films 
was a liquid of concentration $\sim 0.04$~\AA$^{-2}$ which became superfluid at temperatures 
below $1.2$~K. Nevertheless, few years later Boninsegni~\cite{boninsegni} found, using a 
very similar approach to GC and attempting extrapolation to the thermodynamic limit, 
that the hydrogen equilibrium state in K-H$_{2}$ films was a crystal commensurate 
with the underlying lattice of alkali atoms. The superfluid fraction of such 
a commensurate system was equal to zero as reported by Boninsegni.    
Almost simultaneously to the publication of Boninsegni's work~\cite{boninsegni}, Cazorla and 
Boronat presented a ground-state study (i.e., performed at zero temperature) of a two-dimensional 
system composed of Rb atoms and hydrogen molecules~\cite{cazorla04}. By using the diffusion 
Monte Carlo (DMC) method and somewhat more realistic AM-H$_{2}$ potentials than
adopted by GC and Boninsegni, they found that the p-H$_{2}$ ground-state in the Rb-H$_{2}$ film 
was a highly structured liquid with a practically suppressed superfluid fraction 
of $\rho_{s}/\rho = 0.08(2)$. Overall, these theoretical predictions appeared 
to suggest that the embedding of alkali metal atoms on hydrogen matrices was not an
effective strategy to trigger p-H$_{2}$ superfluidity.       

In this work, we report an exhaustive diffusion Monte Carlo (DMC) study of the 
ground-state properties (i.e., energetic, structural and superfluid) of p-H$_{2}$
molecules within a two-dimensional solid matrix of sodium (Na) atoms.
Our main finding is that the p-H$_{2}$ ground-state is a liquid that possesses 
a remarkably large superfluid fraction of $\rho_{s}/\rho = 0.29(2)$. 
The reason behind such a large superfluid 
response lies on the details of the Na-H$_{2}$ interaction, which presents a 
smaller repulsive core as compared to other AM-H$_{2}$ pairwise potentials.  

The organization of this article is as follows. In the next section we provide a brief 
description of the DMC method and the details of our calculations. Next, we present  
our results and compare them with previous computational works. Finally, we summarize 
our main findings in Sec.~\ref{sec:conclusions}. 

\begin{figure*}
\centerline{
\includegraphics[width=1.00\linewidth]{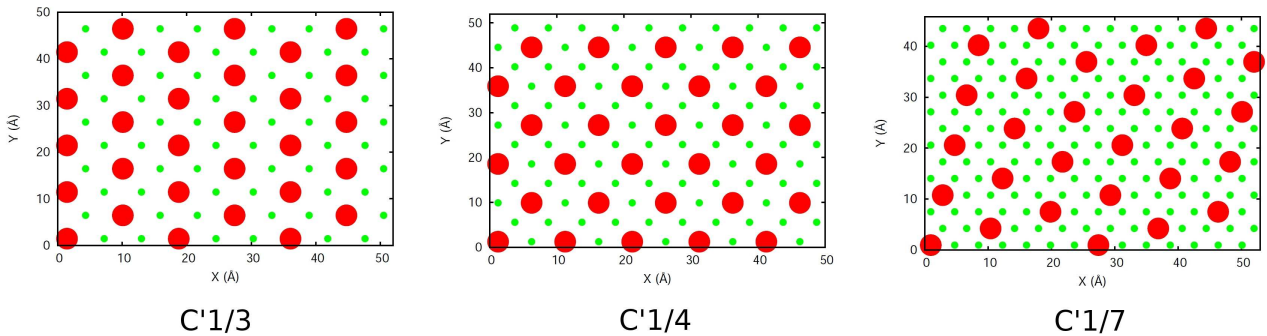}}
\vspace{-0.25cm}
\caption{Representation of the three \emph{pseudo}-commensurate
         crystal structures considered in this work. Big red dots
         represent sodium atoms and small green dots p-H$_{2}$ molecules.
        }
\label{fig0}
\end{figure*}

\section{Methods and Simulation Details}
\label{sec:methods}
The basics of the diffusion Monte Carlo (DMC) method have been
reviewed with detail elsewhere~\cite{hammond94,guardiola98,ceperley86,boronat94}
so here we only comment on the essential ideas.

In the DMC approach, the time-dependent Schr\"odinger equation
of a quantum system of $N$ interacting particles is solved
stochastically by simulating the time evolution of the Green's function
propagator $e^{-\frac{i}{\hbar} \hat{H} t}$ in imaginary time
$\tau \equiv \frac{it}{\hbar}$. For $\tau \to \infty$, sets of configurations (walkers)
$\lbrace {\bf R}_i \equiv {\bf r}_1,\ldots,{\bf r}_N \rbrace$
rendering the probability distribution function
($\Psi_0 \Psi$) are generated, where $\Psi_0$ is the ground-state
wave function of the system and $\Psi$ a guiding wave function (gwf) used 
for importance sampling. Within DMC, \emph{exact} results (i.e.,
subject only to statistical uncertainties) are obtained
for the total ground-state energy and related quantities
in bosonic quantum systems~\cite{aclaration,barnett91,casulleras95}.

We are interested in studying the ground-state of a system
of p-H$_{2}$ molecules immersed in a two-dimensional solid matrix
of Na atoms. We model the Hamiltonian of this system as
\begin{equation} 
H= -\frac{\hbar^2}{2m_{\rm H_{2}}} \sum_{i=1}^{N} \nabla_i^2 + \sum_{i<j}^{N} V_{\rm H_{2}-H_{2}}(r_{ij}) + \sum_{i,k}^{N,n} V_{\rm Na-H_{2}}(R_{ik})~,
\label{eq:hamiltonian}
\end{equation} 
where $m_{\rm H_{2}}$ is the mass of a p-H$_{2}$ molecule, $N$ the number of 
hydrogen molecules, $n$ the number of alkali metal atoms, and $V_{\rm H_{2}-H_{2}}$ and
$V_{\rm Na-H_{2}}$ semi-empirical pairwise potentials describing the
H$_{2}$-H$_{2}$ and Na-H$_{2}$ interactions.  
The internal structure of p-H$_{2}$ molecules has been neglected (i.e., vibrational
and rotational degrees of freedom are disregarded) and the hydrogen-hydrogen molecular 
interactions have been modeled with the standard Silvera-Goldman potential~\cite{silvera}.
The interactions between Na atoms and H$_{2}$ molecules are described with a Lennard-Jones 
potential of the form $V_{LJ} (r) = 4 \epsilon [\left( \sigma/r \right)^{12} - 
\left( \sigma/r \right)^{6}]$, with parameters taken from Ref.~[\onlinecite{Anc}],
namely $\sigma= 4.14$~\AA~ and $\epsilon= 30$~K. The kinetic energy of the 
Na atoms has been also neglected since this is expected to be much smaller 
than the typical energy scale of p-H$_{2}$ molecules (i.e., $10-100$~K). 

It is worth noticing that despite asymptotic DMC results
do not depend on the choice of the guiding wave function (gwf), the algorithmic 
efficiency in DMC runs is influenced by the quality of $\Psi$.
The guiding wave function that we use to describe the present Na-H$_{2}$ system  
contains correlations between the $N$ H$_{2}$ molecules (${\rm f_{2}}(r_{ij})$) and 
the $N$ H$_{2}$ molecules and $n$ alkali metal atoms (${\rm F_{2}}(R_{ij})$).
In the liquid phase, this gwf reads  
\begin{equation}
\label{eq:twfliquid}
\Psi_{L} \left( \mathbf{r}_1, \mathbf{r}_2, \ldots, \mathbf{r}_N \right) = 
\prod_{i<j}^{N}{\rm f_{2}}(r_{ij}) \prod_{i,k}^{N,n}{\rm F_{2}}( R_{ik})
\end{equation}
where two-body correlation factors ${\rm f_{2}}(r)$ and ${\rm F_{2}}(r)$ have been chosen 
of the McMillan form, i.e., $\exp (-\frac{1}{2} (b/r)^{5})$, and $R_{ik}$ is the distance
between the $i$th p-H$_{2}$ molecule and the $k$th alkali atom.   
In order to compute the energy of possible solid \emph{pseudo}-commensurate phases 
(see next section for details), we adopted the guiding wave function
\begin{equation}
\label{eq:twfsolid}
\Psi_{S} \left( \mathbf{r}_1, \mathbf{r}_2, \ldots, \mathbf{r}_N \right) =
\Psi_{L} \prod_{i=1}^{N} {\rm g_{1}}( \xi_{i})
\end{equation}
where one-body factors ${\rm g_{1}}$ are Gaussians, i.e., $\exp (-\frac{1}{2} c r^{2})$,
and $\xi_{i}$ is the distance of the $i$th p-H$_{2}$ molecule to its site in the 
corresponding perfect lattice configuration. 
The value of all variational parameters were determined through subsidiary 
variational Monte Carlo calculations. In the liquid phase, these resulted $b = 3.70$~\AA~  
(${\rm f_{2}}$) and $5.60$~\AA~ (${\rm F_{2}}$), and in the solid phase, 
$b = 3.45$~\AA~ (${\rm f_{2}}$), $5.60$~\AA~ (${\rm F_{2}}$) and $c = 1.22$~\AA$^{-2}$ 
(${\rm g_{1}}$).

In our simulations, both the alkali metal atoms and p-H$_{2}$ molecules are arranged in 
a strictly two-dimensional geometry. Na atoms are considered static and distributed according 
to a triangular lattice of parameter $10$~\AA~. It is worth noticing that such an alkali 
metal geometry is realistic since it has been experimentally observed in Ag(111) 
plates~\cite{Lea,diehl,leatherman}. In order to determine the equation of state and ground-state 
properties of the liquid H$_{2}$ film, we kept the number of alkali atoms fixed to $30$ and 
progressively increased the concentration of p-H$_{2}$ molecules. The typical size of our 
simulation boxes is $50$~\AA$\times 50$~\AA~. The value of the technical parameters 
in the calculations were set to ensure convergence of the total energy per particle to less 
than $0.1$~K/atom. For instance, the mean population of walkers was equal to $400$ and 
the length of the imaginary time-step ($\Delta \tau$) to $5 \cdot 10^{-4}$~K$^{-1}$~.
Statistics were accumulated over $10^{5}$ DMC steps performed after equilibration of the 
system and the approximation used for the short-time Green's function $e^{-\hat{H} \tau}$ 
is exact up to order $(\Delta \tau)^{2}$~\cite{boronat94,chin90}.
It is important to stress that by using the same DMC method we 
have been able to reproduce in previous works the experimental equation of state of 
archetypal quantum solids like $^{4}$He, H$_{2}$, LiH, and Ne~\cite{cazorla08,cazorla08a,boronat04,cazorla08c,cazorla10,cazorla09,gordillo11,lutsyshyn10,cazorla05}.   

\begin{figure}
\centerline
        {\includegraphics[width=0.9\linewidth]{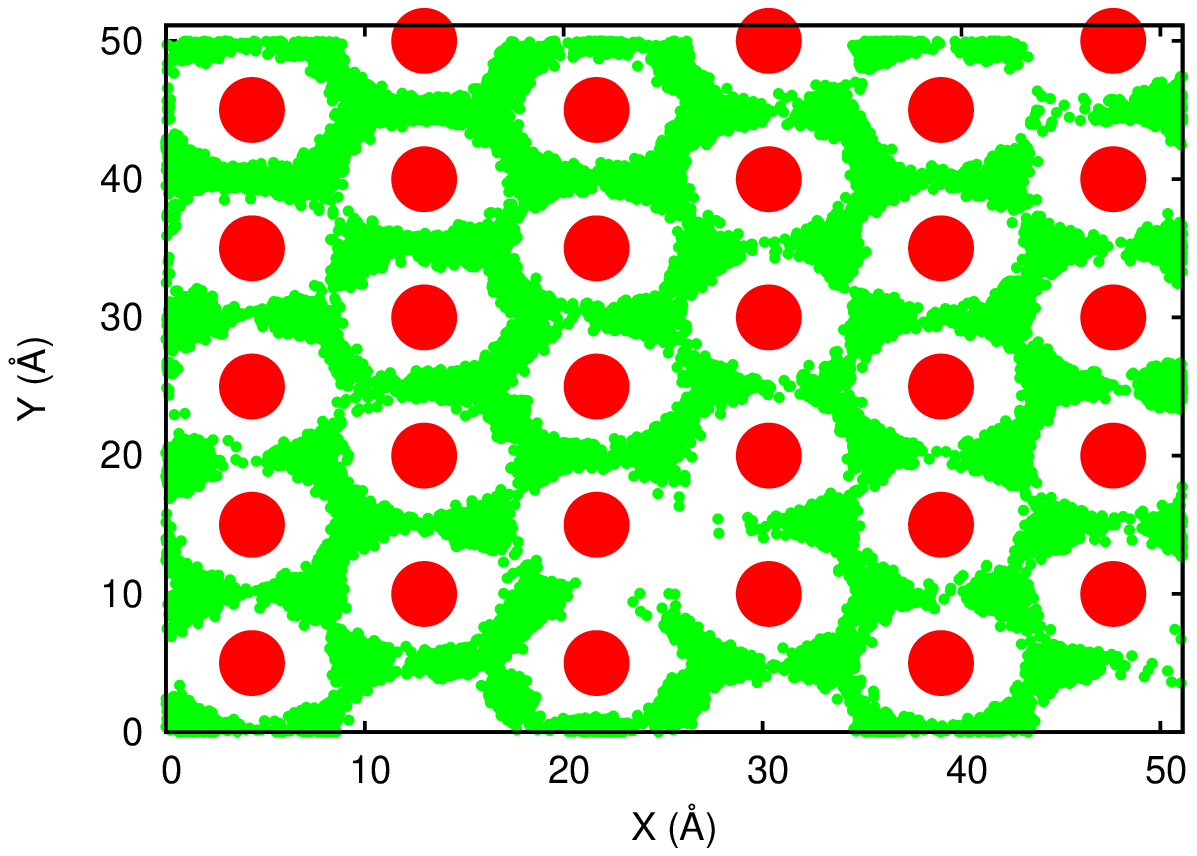}}%
        {\includegraphics[width=0.9\linewidth]{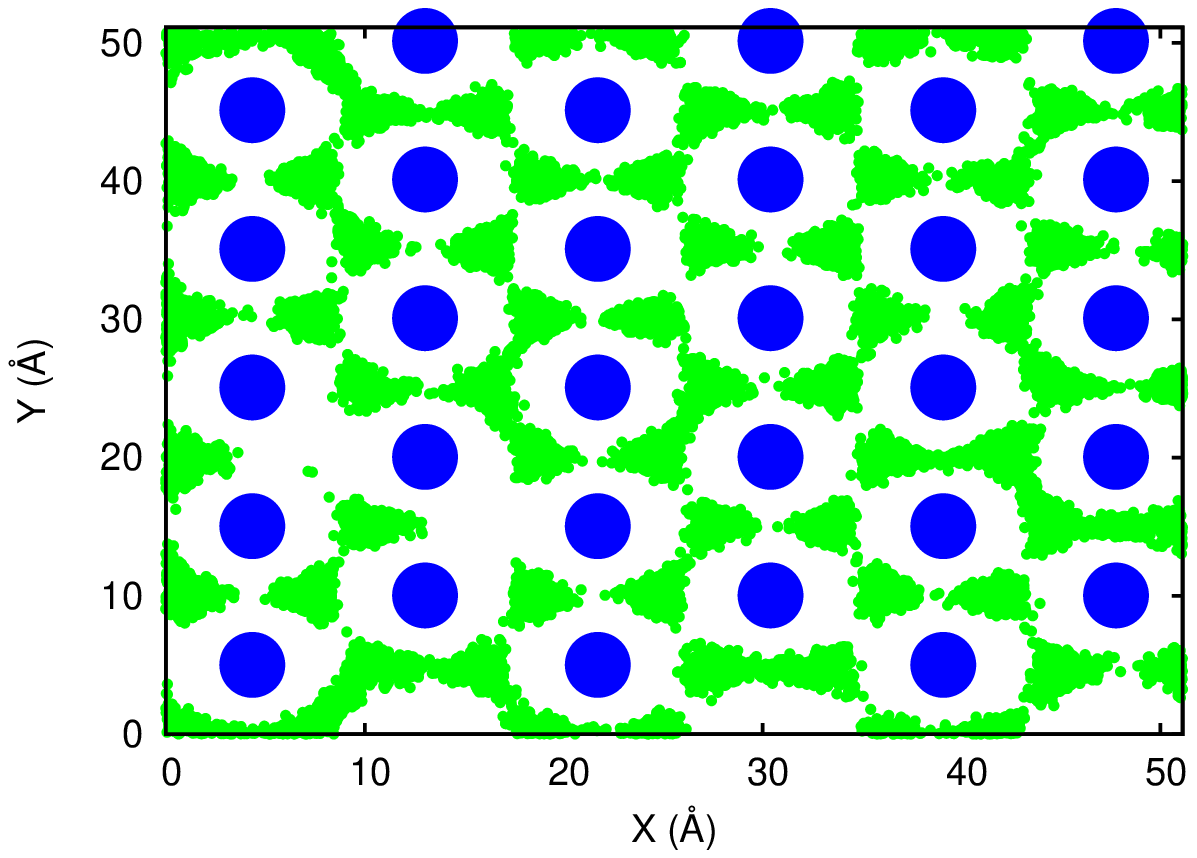}}%
\caption{Snapshot of the probability density of two-dimensional p-H$_{2}$
(green dots) calculated in the liquid Na-H$_{2}$ (Na = red dots, top) and Rb-H$_{2}$
(Rb = blue dots, bottom) films at equilibrium (i.e., $0.038$ and $0.023$~\AA$^{-2}$,
respectively) and zero temperature.}
\label{fig1}
\end{figure}

\section{Results and Discussion}
\label{sec:results}

\begin{table}
\begin{center}
\begin{tabular}{ c c c c }
\hline
\hline
 $ $ & $ $ & $ $ & $ $ \\
$\rho$~(\AA$^{-2}$) & $ E/N$  &  $\langle V\rangle/N $ & $\langle T\rangle/N $ \\
 $ $ & $ $ & $ $ & $ $ \\
\hline
 $ $ & $ $ & $ $ & $ $ \\
$0.029$ & $-45.04(4)$ & $-63.65(7)$ & $18.61(7)$ \\
$0.031$ & $-45.66(3)$ & $-65.52(5)$ & $19.86(5)$ \\
$0.033$ & $-46.24(2)$ & $-67.02(5)$ & $20.78(5)$ \\
$0.036$ & $-46.93(2)$ & $-69.36(7)$ & $22.43(7)$ \\
$0.038$ & $-47.13(2)$ & $-71.57(7)$ & $24.44(7)$ \\
$0.040$ & $-46.86(3)$ & $-72.95(5)$ & $26.09(5)$ \\
 $ $ & $ $ & $ $ & $ $ \\
\hline
\hline
\end{tabular}
\end{center}
\caption{Total ($E$), potential ($V$) and kinetic ($T$) energies per p-H$_{2}$ molecule 
calculated near equilibrium in the liquid H$_{2}$ system expressed in units
of K.}
\end{table}

\begin{table}
\begin{center}
\begin{tabular}{ c c c c c }
\hline
\hline
 $ $ & $ $ & $ $ & $ $ & $  $ \\
 ${\rm Phase} $ &  $\rho$~(\AA$^{-2}$) & $ E/N$ &  $\langle V\rangle/N $ & $\langle T\rangle/N $ \\
 $ $ & $ $ & $ $ & $ $ & $  $ \\
\hline
 $ $ & $ $ & $ $ & $ $ & $  $ \\
 $C'_{1/3} $ & $0.023$ & $-41.56(2)$ & $-53.74(3)$ & $12.18(3)$ \\
 $C'_{1/4} $ & $0.035$ & $-46.28(2)$ & $-67.76(5)$ & $21.48(5)$ \\
 $C'_{1/7} $ & $0.069$ & $ 58.47(2)$ & $-22.11(4)$ & $80.58(4)$ \\
 $ $ & $ $ & $ $ & $ $ & $  $ \\
\hline
\hline
\end{tabular}
\end{center}
\caption{Total ($E$), potential ($V$) and kinetic ($T$) energies per p-H$_{2}$ molecule 
         calculated in three different \emph{pseudo}-commensurate solid H$_{2}$ systems
         expressed in units of K.}
\end{table}

Let us to start by presenting the energy results obtained in the 
liquid Na-H$_{2}$ system. The correspoding total, potential and kinetic 
energies per hydrogen molecule expressed as a function of density 
are enclosed in Table~I (potential energies were obtained with the pure 
estimator technique hence all the reported energies are \emph{exact},
i.e., subject to statistical uncertainty only~\cite{barnett91,casulleras95}).
The ground-state energy and equilibrium density of the liquid film,
$e_0$ and $\rho_0$, were determined by fitting the polynomial curve 
\begin{equation}
\label{eq:efit}
e(\rho) = e_0 + B\left(\frac{\rho-\rho_0}{\rho_0}\right)^2+C\left(\frac{\rho-\rho_0}{\rho_0}\right)^3
\end{equation}
to the calculated total energies.
The resulting optimal parameter values are $B = 86.16$~K, $C = 221.69$~K, 
$\rho_{0} = 0.038$~\AA$^{-2}$ and $e_{0} = -47.13$~K. 
We note that the liquid ground-state energy and equilibrium density are significantly 
different from those computed in the pure two-dimensional p-H$_{2}$ crystal, 
namely $-23.41$~K and $0.067$~\AA$^{-2}$~\cite{cazorla08}. 
These large total energy and equilibrium density differences have their origin in 
the potential energy gain and steric effects deriving from the presence of sodium atoms. 
Also, we observe that the equilibrium density of the liquid H$_{2}$ film 
is appreciably larger than the calculated in the analogous Rb-H$_{2}$ system, 
namely $0.023$~\AA$^{-2}$~\cite{cazorla04} (e.g., the ratio among the number of 
p-H$_{2}$ molecules and alkali metal atoms are $10/3$ and $2/1$, 
respectively). The cause for the large equilibrium density in the Na case,
as compared to that of the Rb system, is related to the decrease of the core size
of the AM-H$_{2}$ interaction (i.e., $\sigma = 4.54$~\AA~ in the Rb-H$_{2}$ case), 
which makes the surface available to p-H$_{2}$ molecules larger (we note that
the depth of the potential wells, $\epsilon$, in both Na-H$_{2}$ and Rb-H$_{2}$ 
interactions are very similar, i.e., $30$ and $28$~K respectively~\cite{Anc}).

Regarding the stabilization of possible \emph{pseudo}-commensurate solid phases, we
investigated the three crystal structures shown in Fig.~\ref{fig0}. We refer to them 
as \emph{pseudo}-commensurate phases because in order to fully fulfil commensurability 
some p-H$_{2}$ molecules should be located at the same $x-y$ positions than alkali atoms. 
Since the system considered in the present study is strictly two-dimensional, this positional 
coincidence is energetically forbidden. Therefore, we started by generating 
the exact $C_{1/3}$, $C_{1/4}$ and $C_{1/7}$ commensurate structures (where the 
subscripts indicate the relative population of alkali and hydrogen species) and then 
removed by hand the p-H$_{2}$ molecules located at the same positions than sodium atoms
(hence the prime in our notation). In Table~II, we enclose the energy results obtained
for those \emph{pseudo}-commensurate structures. It is found that of the three cases
considered $C'_{1/4}$ is by far the system with the lowest energy. 
Interestingly, the density of the $C'_{1/4}$ phase (i.e., $0.035$~\AA$^{-2}$) 
is very close to the equilibrium density found in the analogous liquid  
system (i.e., $0.038$~\AA$^{-2}$). Moreover, 
from a structural point of view the \emph{pseudo}-commensurate $C'_{1/4}$ phase is very 
similar to the equilibrium state predicted by Boninsegni in K-H$_{2}$ films at low
temperatures (see Fig.~3 in Ref.~\cite{boninsegni}). Nevertheless, the total energy 
per particle of the $C'_{1/4}$ phase is about $0.9$~K larger than the energy 
of the corresponding fluid at equilibrium and thus, according to our calculations, the 
ground-state of the Na-H$_{2}$ system is a liquid. In view of these findings, we will 
concentrate on the description of the liquid H$_{2}$ system in the remainder of this 
article.      

In Fig.~\ref{fig1}, we show a snapshot of the probability density 
calculated for the ground-state of the Na-H$_{2}$ system. There it is observed 
that hydrogen molecules can access a large portion of the surface left 
between Na atoms by diffusing through honeycomb-like pathways created around 
the alkali metal centers. This situation is eminently different from the one
observed in Rb-H$_{2}$ films, where a highly structured liquid is found to 
be the ground-state (see Ref.~\cite{cazorla04} and Fig.~\ref{fig1}). In this last case,
most of p-H$_{2}$ molecules are localized within the interior of the triangles 
formed by Rb atoms and the connectivity between high-density p-H$_{2}$ regions 
is rather low. The probability density differences observed between 
Na-H$_{2}$ and Rb-H$_{2}$ systems again can be understood in terms of the  
core lengths of the corresponding AM-H$_{2}$ interactions.

\begin{figure}
\centerline
{\includegraphics[width=1.00\linewidth]{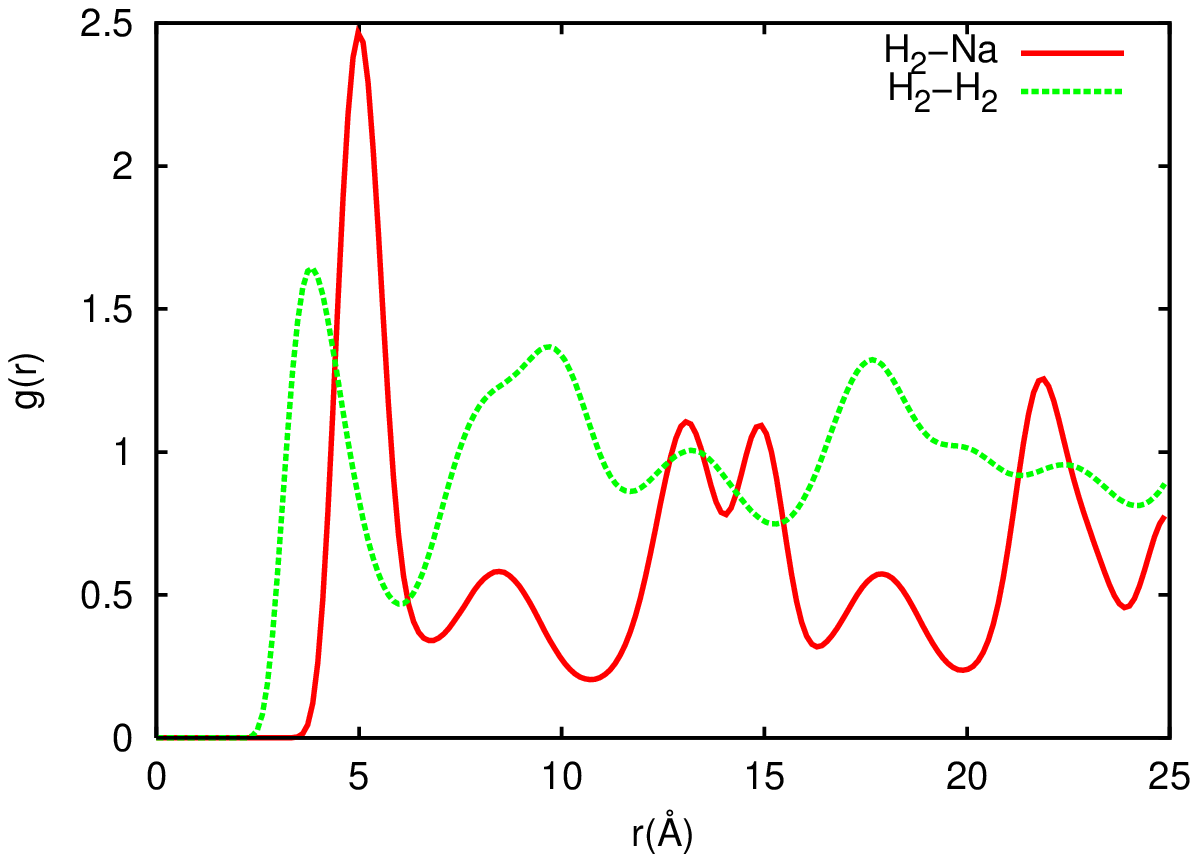}}%
{\includegraphics[width=1.00\linewidth]{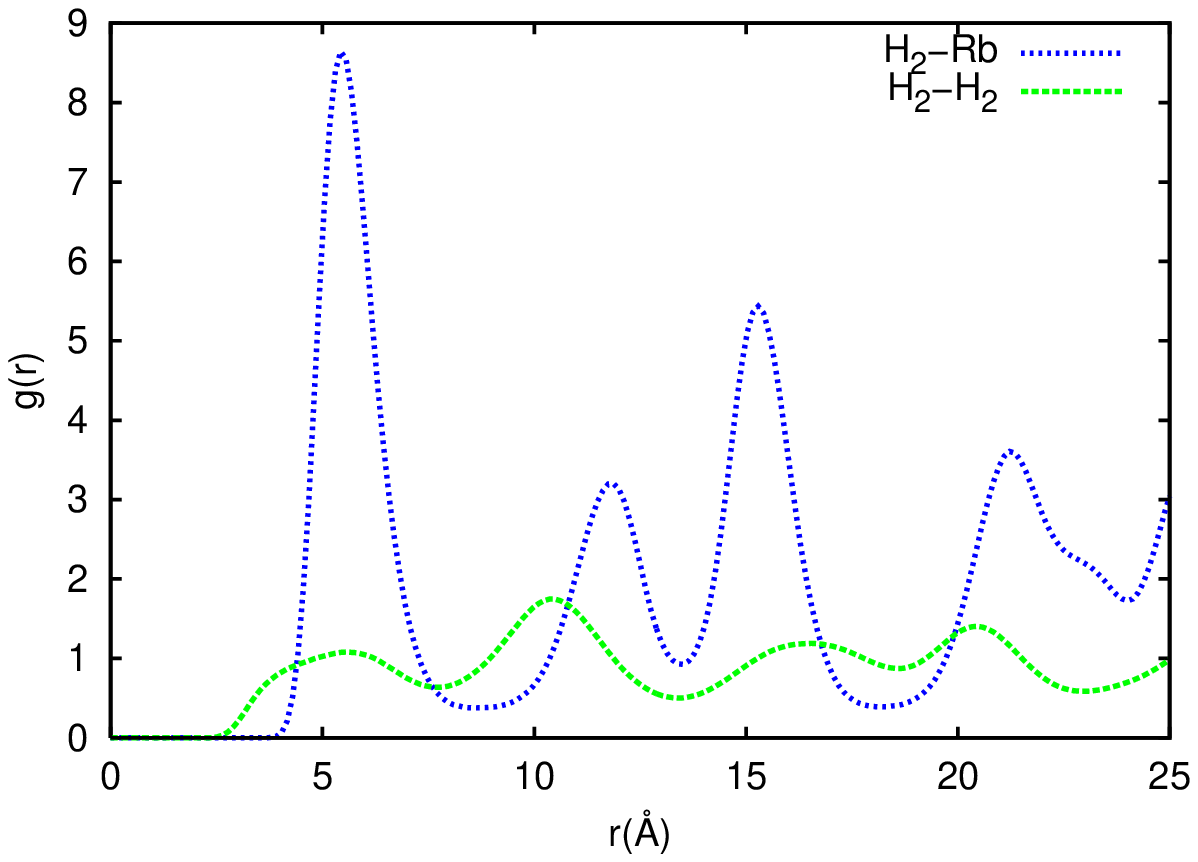}}%
\caption{Pair-radial distribution functions obtained
         in the liquid Na-H$_{2}$ (top) and Rb-H$_{2}$ (bottom) 
         films at their corresponding equilibrium densities.}
\label{fig2}
\end{figure}

The enhanced delocalization of p-H$_{2}$ molecules in the Na-based film 
can be also deduced from the shape of the calculated pair-radial distribution functions. 
In Fig.~\ref{fig2}, one can observe that the peaks of the crossed $g_{\rm Na-H_{2}}$ 
distribution function are less sharp than those obtained in the analogous Rb-H$_{2}$ 
system ($g(r)$ results from Ref.~\cite{cazorla04} have been included in the plot for 
comparison purposes). 
Also, the first peak of the $g_{\rm H_{2}-H_{2}}$ function centered at $r = 3.9$~\AA~ is 
a global maximum and does not coincide with the position of the first $g_{\rm Na-H_{2}}$ 
peak found at $r = 5.1$~\AA~, so implying a high concentration of p-H$_{2}$ molecules.
These last features are in opposition to what is observed in the Rb-H$_{2}$ film, where 
the first and second $g_{\rm H_{2}-H_{2}}$ peaks centered at $5.2$ and $10.3$~\AA~ can be 
ascribed to the hexagonal-like pattern that results from filling the triangles 
formed by Rb atoms with one p-H$_{2}$ molecule. 

\begin{figure}
\centerline{
\includegraphics[width=1.00\linewidth]{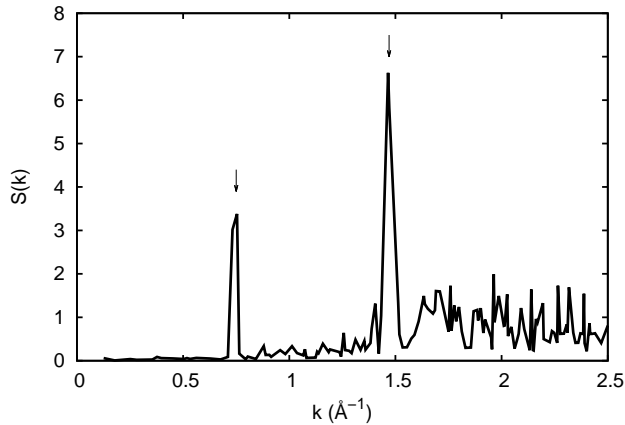}}
\caption{Molecular hydrogen structure factor obtained
        in the liquid Na-H$_{2}$ system at the equilibrium density.
        Peaks ascribed to the periodicity of the underlying  
        sodium film are indicated with arrows.}
\label{fig3}
\end{figure}

\begin{figure}
\centerline{
\includegraphics[width=1.00\linewidth]{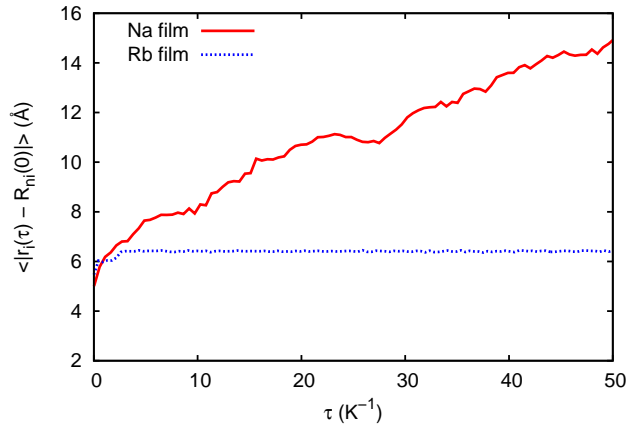}}
\caption{Averaged mean distance of p-H$_{2}$ molecules
         to the nearest Na and Rb atoms at the 
         start of the simulations carried out at the respective 
         equilibrium densities.}
\label{fig4}
\end{figure}

In order to better assess the structure of the p-H$_{2}$ molecules in the Na 
film, we calculated the corresponding structure factor, $S({\bf k})$,  
shown in Fig.~\ref{fig3}. There, one can observe the presence of two sharp peaks centered 
at reciprocal lattice vectors that essentially coincide with the periodicity imposed 
by the triangular Na lattice (i.e., at ${\bf k} = 0.75$ and $1.5$~\AA$^{-1}$).
However, no other large scattering peaks signalizing the appearance of a solid or glassy 
state are seen in the figure. The p-H$_{2}$ film, therefore, appears to be a  
fluid. As an additional test, we also monitored the average distance that the p-H$_{2}$ 
molecules move away from the Na atoms which at the start of the simulation are closest 
to them. We plot this quantity as a function of imaginary time in Fig.~\ref{fig4}. 
As it can be appreciated, function $\Delta r (\tau) = \langle |{\bf r}_{i}(\tau) 
- {\bf R}_{ni}(0)| \rangle$ monotonically increases with $\tau$ reproducing so the typical 
profile obtained in bulk fluids (i.e., is roughly linear). Assuredly, then, the 
simulated p-H$_{2}$ system remains in a liquid phase. For comparison purposes, we include 
also in Fig.~\ref{fig4} the diffusion profile obtained in the equivalent Rb-H$_{2}$ system
under equilibrium conditions. In this last case, the mobility of the hydrogen molecules is 
practically suppressed as shown by the computed $d \Delta r /d \tau \sim 0$ 
slope.  

\begin{figure}
\centerline{
\includegraphics[width=1.00\linewidth]{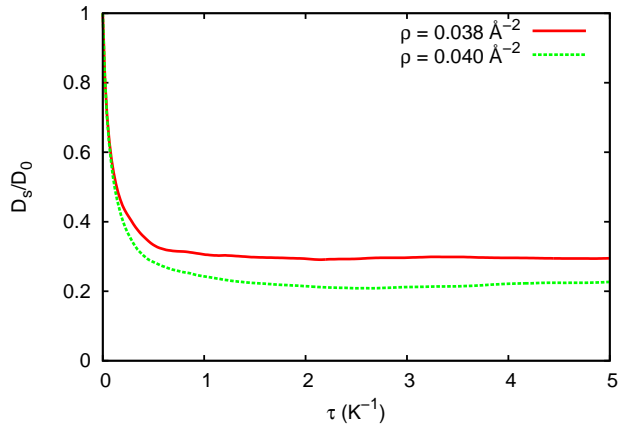}}
\caption{Diffusion Monte Carlo estimation of the p-H$_{2}$ superfluid density 
         in Na films at different densities.}
\label{fig5}
\end{figure}

In order to complete our description of the Na-H$_{2}$ film, we estimated 
the superfluid fraction, $\rho_{s}/\rho$, of the p-H$_{2}$ sub-system.
The superfluid fraction of a bosonic system can be estimated within the DMC formalism
by extending to zero temperature the winding-number technique employed in PIMC
calculations~[\onlinecite{super}].
In two dimensions, $\rho_{s}/\rho$ is estimated as the ratio of two diffusion
constants, $D_{s}/D_{0}$, in the imaginary time limit $\tau \rightarrow \infty$,
where $D_{0} = \hbar^{2}/(2m_{\rm H_{2}})$ and
\begin{equation}
D_{s}= \lim_{\tau \rightarrow \infty} \frac{N}{4 \tau} \, \frac{\int d \mathbf{R}
f(\mathbf{R},\tau) \left[ \mathbf{R}_{\rm{CM}}(\tau)-\mathbf{R}_{\rm{CM}}(0) \right]^2}
{\int d \mathbf{R}
f(\mathbf{R},\tau)}~,
\label{eq:super2D}
\end{equation}
with $\mathbf{R}_{\rm{CM}}=(1/N) \sum_{i=1}^{N} \mathbf{r}_i$ being the position
of the p-H$_{2}$ center of mass. In Fig.~\ref{fig5}, we plot the $D_{s}/D_{0}$ function 
computed at two different densities and expressed as a function of imaginary 
time. From the $D_{s}/D_{0}$ asymptote we estimate the superfluid fraction of the 
hydrogen liquid at equilibrium to be $0.29(2)$, a quite large value. It is worth recalling 
that the superfluid fraction computed in the equivalent Rb-H$_{2}$ system was much smaller, 
namely $\rho_{s}/\rho = 0.08(2)$~\cite{cazorla04}. This significant $\rho_{s}/\rho$ 
difference is a direct consequence of the increase in the p-H$_{2}$ concentration  
at equilibrium, which in turn depends on the form of the AM-H$_{2}$ interaction. 
Furthermore, we calculated the superfluid response of the Na-H$_{2}$ film at a density 
slightly above the equilibrium point (see Fig.~\ref{fig5}) and found that the value 
of the $D_{s}/D_{0}$ asymptote decreases (i.e., $0.23(2)$ at $\rho = 0.040$~\AA$^{-2}$). 
This last finding points out to a strong dependence of the superfluid fraction
on $\rho$ due to the effect of excluding surface produced by the presence
of static Na atoms. We note that the finite size of the simulation box could induce 
some dependence of the superfluid fraction on the number of particles. In order to 
reduce this effect, however, we worked out with a rather large simulation box 
of typical size $50$~\AA$\times 50$~\AA~ (i.e., as large as the one employed in 
Ref.~\cite{boninsegni} where suppresion of p-H$_{2}$ superfludity in a K film 
was predicted).

In the light of our energy, structural and superfluid fraction results obtained 
in alkali-H$_{2}$ films (see work~\cite{cazorla04}) it may be concluded that 
(i)~the softer the repulsive core of the crossed AM-H$_{2}$ interaction is, the larger 
the p-H$_{2}$ equilibrium density and superfluid fraction result, and (ii)~the superfluid 
response of two-dimensional p-H$_{2}$ films strongly depends on density.

\section{Conclusions}
\label{sec:conclusions}
Summarizing, we have performed an exhaustive diffusion Monte Carlo study
of the energetic, structural and superfluid properties of a p-H$_{2}$ two-dimensional
system in which Na atoms have been embedded forming a triangular lattice. 
The main motivation of this computational study was to investigate whether hydrogen 
crystallization could be prevented in two dimensions and, if so, to estimate the superfluid 
response of the corresponding p-H$_{2}$ sub-system at zero-temperature. 
We have found, in contrast to previous computational works considering other alkali 
metal species and AM-H$_{2}$ potentials, that the p-H$_{2}$ ground-state in the 
Na film is a liquid that possesses a remarkably large superfluid fraction 
(i.e., $\rho_{s}/\rho = 0.29(2)$). 
The principal reason behind the stabilization of this fluid relies on the 
fact that Na-H$_{2}$ interactions are less attractive than H$_{2}$-H$_{2}$ 
and thus a significant reduction of the hydrogen equilibrium density occurs. 
Also, we have found that the energetic, structural and superfluid properties 
of p-H$_{2}$ films strongly depend on density.

Importantly, we note that small variations of the crossed AM-H$_{2}$ 
potential parameters may lead to appreaciable differences on the computed 
p-H$_{2}$ properties. Therefore, since there are few potentials in the literature 
which describe the interactions between alkali metal atoms and p-H$_{2}$ molecules 
accurately, and those which have been reported probably are not too versatile, we must 
be cautious at our conclusions. More realistic and transferable alkali-H$_{2}$ 
potentials than currently available are urgently needed to provide decisive hints 
in the quest for realizing p-H$_{2}$ superfluidity. Nevertheless, in view of the great 
fundamental interest of possible p-H$_{2}$ superfluidity we strongly encourage 
experimental realizations of molecular hydrogen films adsorbed on Na substrates.    

\acknowledgments
This work was supported by MICINN-Spain [Grants No.~MAT2010-18113, 
No.~CSD2007-00041, and FIS2011-25275], Generalitat de Catalunya [Grant
No.~2009SGR-1003], and the CSIC JAE-doc (C.C.) program.


\begin{thebibliography}{30}

\bibitem{vanstraaten} J. van Straaten, R. J. Wijngaarden and I. F. Silvera,
                      Phys. Rev. Lett. {\bf 48}, 97 (1982). 
\bibitem{boronat12} O. N. Osychenko, R. Rota, and J. Boronat, Phys. Rev. B {\bf 85}, 
                    224513 (2012).
\bibitem{seidel} G. M. Seidel, H. J. Maris, F. I. B. Williams and J. G. Cardon, 
                 Phys. Rev. Lett. {\bf 56}, 2380 (1986).
\bibitem{maris} H. J. Maris, G. M. Seidel and F. I. B. Williams, 
                Phys. Rev. B {\bf 36}, 6799 (1987).
\bibitem{dekinder} J. De Kinder, A. Bouwen and D. Schoemaker, 
                   Phys. Rev. B \textbf{52}, 15782 (1995).
\bibitem{brewer} D. F. Brewer, J. C. N. Rajendra and A. L. Thomson, 
                 J. Low Temp. Phys. \textbf{101}, 317 (1995).
\bibitem{schindler} M. Schindler, A. Dertinger, Y. Kondo and F. Pobell, 
                    Phys. Rev. B \textbf{53}, 11451 (1996).
\bibitem{sokol} P. E. Sokol, R. T. Azuah, M. R. Gibbs and S. M. Bennington, 
                J. Low Temp. Phys. \textbf{103}, 23 (1996).
\bibitem{liu} F. C. Liu, Y. M. Liu, and O. E. Vilches,
              Phys. Rev. B \textbf{51}, 2848 (1995).
\bibitem{grebenev} S. Grevenev, B. Sartakov, J. P. Toennius and A. F. Vilesov,
                   Science \textbf{289}, 1532 (2000).
\bibitem{boro} M. C. Gordillo, J. Boronat, and J. Casulleras,
               Phys. Rev. Lett. \textbf{85}, 2348 (2000).
\bibitem{gordillo01} M. C. Gordillo and D. M. Ceperley,
                     Phys. Rev. B \textbf{65}, 174527 (2001).
\bibitem{drops} P. Sindzingre, D. M. Ceperley, and M. L. Klein,
                Phys. Rev. Lett. \textbf{67}, 1871 (1991).
\bibitem{gordillo} M. C. Gordillo and D. M. Ceperley,
                   Phys. Rev. Lett. \textbf{79}, 3010 (1997).
\bibitem{boninsegni} M. Boninsegni, New Journal of Physics \textbf{7}, 78 (2005). 
\bibitem{cazorla04} C. Cazorla and J. Boronat, 
                    J. of Low. Temp. Phys. \textbf{134}, 43 (2004).
\bibitem{hammond94} B. L. Hammond, W. A. Lester, and Jr. P. J. Reynolds
                    in \textit{Monte Carlo Methods in Ab Initio Quantum Chemistry},
                    World Scientific, Singapore (1994).
\bibitem{guardiola98} R. Guardiola, Lecture Notes in Physics \textbf{510}, 269 (1998).
\bibitem{ceperley86} D. M. Ceperley and M. H. Kalos in \textit{Monte Carlo methods in
                     statistics physics}, Springer-Verlag, Berlin (1986).
\bibitem{boronat94} J. Boronat and J. Casulleras, Phys. Rev. B \textbf{49}, 8920 (1994).
\bibitem{aclaration} By related quantities is meant the expected value of
                     operators $\hat{A}$ that commute with the Hamiltonian, namely
                     $[\hat{A}, \hat{H}] = 0$~. It is also possible to
                     obtain virtually exact results for
                     $[\hat{A}, \hat{H}] \neq 0$ operators
                     by using forward walking based techniques (see Refs.~[\onlinecite{barnett91}]
                     and [\onlinecite{casulleras95}]).
\bibitem{barnett91} R. Barnett, P. Reynolds, and W. A. Lester Jr., J. Comput. Phys.
                    \textbf{96}, 258 (1991).
\bibitem{casulleras95} J. Casulleras and J. Boronat, Phys. Rev. B \textbf{52}, 3654 (1995).
\bibitem{silvera} I. F. Silvera and V. V. Goldman,
                  J. Chem. Phys. \textbf{69}, 4209 (1978).
\bibitem{Anc} F. Ancilotto, E. Cheng, M. W. Cole, and F. Toigo,
              Z. Phys. B \textbf{98}, 323 (1995).
\bibitem{Lea} G. S. Leatherman and R. D. Diehl,
              Phys. Rev. B \textbf{53}, 4939 (1996).
\bibitem{diehl} R. D. Diehl, \emph{private communication}.
\bibitem{leatherman} Gerald S. Leatherman, \textit{Ph.D. thesis} 
                     (Penn State University, 1996).
\bibitem{chin90} S. A. Chin, Phys. Rev. A \textbf{42}, 6991 (1990).
\bibitem{cazorla08} C. Cazorla and J. Boronat, 
                    Phys. Rev. B \textbf{78}, 134509 (2008).
\bibitem{cazorla08a} C. Cazorla and J. Boronat, J. Phys.: Condens. Matter \textbf{20}, 015223 (2008).
\bibitem{boronat04} J. Boronat, C. Cazorla, D. Colognesi, and M. Zoppi, Phys. Rev. B \textbf{69},
                    174302 (2004).
\bibitem{cazorla08c} C. Cazorla and J. Boronat, Phys. Rev. B \textbf{77}, 024310 (2008).
\bibitem{cazorla10}  C. Cazorla, G. Astrakharchick, J. Casulleras, and J. Boronat,
                     J. Phys.: Condens. Matter \textbf{22}, 165402 (2010).
\bibitem{cazorla09} C. Cazorla, G. Astrakharchick, J. Casulleras, and J. Boronat,
                     New Journal of Phys. \textbf{11}, 013047 (2009).
\bibitem{gordillo11} M. C. Gordillo, C. Cazorla, and J. Boronat,
                     Phys. Rev. B \textbf{83}, 121406(R) (2011).
\bibitem{lutsyshyn10} Y. Lutsyshyn, C. Cazorla, G. E. Astrakharchik, and J. Boronat,
                      Phys. Rev. B \textbf{82}, 180506(R) (2010)
\bibitem{cazorla05} C. Cazorla and J. Boronat, J. Low Temp. Phys. \textbf{139}, 645 (2005).
\bibitem{super} S. Zhang, N. Kawashima, J. Carlson, and J. E. Gubernatis,
                Phys. Rev. Lett. \textbf{74}, 1500 (1995).

\end{thebibliography}
\end{document}